\documentclass[10pt,aps,pra,twocolumn,floatfix]{revtex4-2}
\usepackage{epsfig}
\usepackage{graphicx}
\usepackage{dcolumn}
\usepackage{amsthm,amsmath,amsfonts}
\usepackage{mathtools}
\usepackage{comment}
\usepackage{xcolor}
\usepackage{orcidlink}
\usepackage{rotating}
\usepackage{cancel}
\usepackage[normalem]{ulem}
\usepackage{bm}
\usepackage{latexsym}
\usepackage{verbatim}
\usepackage{stmaryrd}
\usepackage{amssymb}
\usepackage{braket}
\usepackage{booktabs}
\usepackage{lineno}
\usepackage{subcaption}
\usepackage{multirow}

\begin{document}
\title{Atomic Transition Rates on a Quantum Annealer}
\author{$^{a,b}$Boni Paul\orcidlink{0009-0007-8355-6571}}
\email{boni.paul.71@tcgcrest.org}
\author{$^a$Subimal Deb\orcidlink{0000-0001-7697-1266}}
\email{subimal.deb@tcgcrest.org}
\author{$^c$Per Jönsson\orcidlink{0000-0001-6818-9637}}
\email{per.jonsson@mau.se}
\author{$^c$Jörgen Ekman}
\email{jorgen.ekman@mau.se}
\author{$^{a,d}$Bhanu Pratap Das\orcidlink{0000-0002-3766-6979}}
\email{Bhanu.das@tcgcrest.org}
\affiliation{$^a$Centre for Quantum Engineering Research and Education (CQuERE), TCG CREST, Kolkata, India}
\affiliation{$^b$Department of Physics, Indian Institute of Technology Tirupati, Yerpedu, Andhra Pradesh 517619, India.}
\affiliation{$^c$Department of Materials Science and Applied Mathematics, Malmö University, SE-20506 Malmö, Sweden}
\affiliation{$^d$Department of Physics, School of Science, Tokyo Institute of Technology, Ookayama, Meguro-ku, Tokyo 152-8550, Japan}

\begin{abstract}
Electric-dipole (E1) transition rates underlie applications across many areas of
physics and demand more accurate wavefunctions than
{ground-state energies do.}
We report the first calculation of E1 transition rates on a D-Wave quantum
annealer, for neutral Li, Li-like Be, neutral Na, and Na-like Mg. We use a
zooming-and-sigma-annealing scheme with floating-point encoding within the
Quantum Annealer Eigensolver (QAE) algorithm, which yields ground-state
eigenvectors accurate to the chosen three decimal places. Treating the lowest
even-parity ($^2S_{1/2}$) and odd-parity ($^2P_{1/2}$, $^2P_{3/2}$) states as separate
angular-symmetry sectors gives QAE access to excited atomic configurations for
transition-rate calculations. Our results reproduce relativistic configuration-interaction
values obtained classically to within numerical precision,
demonstrating that the QAE framework extends to a broader class of atomic
properties on near-term quantum hardware.
\end{abstract}
\maketitle

Quantum annealing (QA) \cite{PhysRevE.58.5355, Hauke_2020}, a heuristic
optimization method, has been applied to broad areas of practical relevance
\cite{PhysRevE.70.057701, 10.1098/rsta.2021.0417, Yarkoni_2022} owing to its
ability to efficiently solve combinatorial optimization problems by exploiting
quantum effects such as tunneling and superposition. Recent experiments on
available quantum annealers have even claimed a quantum advantage
\cite{doi:10.1126/science.ado6285} by outperforming classical simulations for
some specific state-of-the-art many-body problems. The Quantum Annealer
Eigensolver (QAE) \cite{Teplukhin2019}, a quantum--classical hybrid algorithm,
has been demonstrated to yield high-accuracy results for various physics problems
\cite{kumar2024computation, D0CP04272B, Zade2025, Teplukhin2020, Teplukhin2021,
Teplukhin2019, PhysRevA.106.052605, sahoo2025applicationquantumannealingcomputation,
paul2026applicationsquantumannealingmagnetic}
{such as scattering cross sections, vibrational spectra, and
electric dipole moments, among others.}
In this work, we apply a modified version of QAE to compute an important atomic
property, the electric-dipole (E1) transition rate. Atomic transition rate
calculations are essential for many physics applications, particularly in plasma
diagnostics and the interpretation of laboratory and astrophysical spectra
\cite{ref1, Pradhan2011AtomicAA}, where they are used to determine element
abundances, temperatures, and electron densities, as well as in the development
of atomic clocks and tests of fundamental physics \cite{chakraborty2023high,
PhysRevD.103.L111303}.

Accurate calculation of atomic properties beyond ground-state energies remains
challenging owing to electron correlation effects. These effects manifest
differently when computing the same property across different systems
\cite{VajedSamii1982} or different properties for the same system
\cite{PhysRevD.103.L111303}, as their magnitude and dominant contributions can
vary substantially depending on the specific observable under consideration
\cite{PhysRevD.103.L111303}. Computing the E1 transition rate illustrates these
challenges, as it requires precise determination of several states. This motivates our investigation of whether quantum annealers can capture
the subtle yet essential many-body effects encoded through the Hamiltonian matrix in the basis of configuration state functions (CSFs). The QAE
algorithm inherently yields the lowest-energy eigenstate within a given symmetry
sector. To access excited configurations relevant for E1 transitions, we treat
the even-parity ($s_{1/2}$) and the odd-parity ($p_{1/2}$ and $p_{3/2}$) states
as independent angular-symmetry sectors. This enables the computation of the
corresponding low-lying states and, consequently, the evaluation of E1 transition
rates within the QAE framework.

As a first step toward computing fundamental atomic properties of practical
relevance, we employ the QAE algorithm on a D-Wave quantum annealer to compute
the E1 transition rates in the Babushkin gauge for four atomic systems: neutral $\mathrm{Li}$, Li-like
$\mathrm{Be}$, neutral $\mathrm{Na}$, and Na-like $\mathrm{Mg}$.
{We focus on the lowest states of two angular-momentum symmetries:
the even-parity $^2S_{1/2}$ ground state and the odd-parity $^2P_{1/2}$ and
$^2P_{3/2}$ excited states.}
Our study also demonstrates that this modified QAE method can be extended to
calculate a wide range of other atomic properties, thereby broadening the
applicability of the QAE framework to atoms and molecules. To validate our QAE
values, we benchmark our D-Wave results against classical calculations based on
the configuration-interaction (CI) method employing multiconfiguration
Dirac--Hartree--Fock (MCDHF) orbitals, as implemented in the General-Purpose
Relativistic Atomic Structure Package (GRASP) \cite{atoms11010007,
FroeseFischer2019GRASP}.

\begin{figure*}[htbp!]
    \centering
        \includegraphics[trim=0 16.5cm 0 0, width=\linewidth]{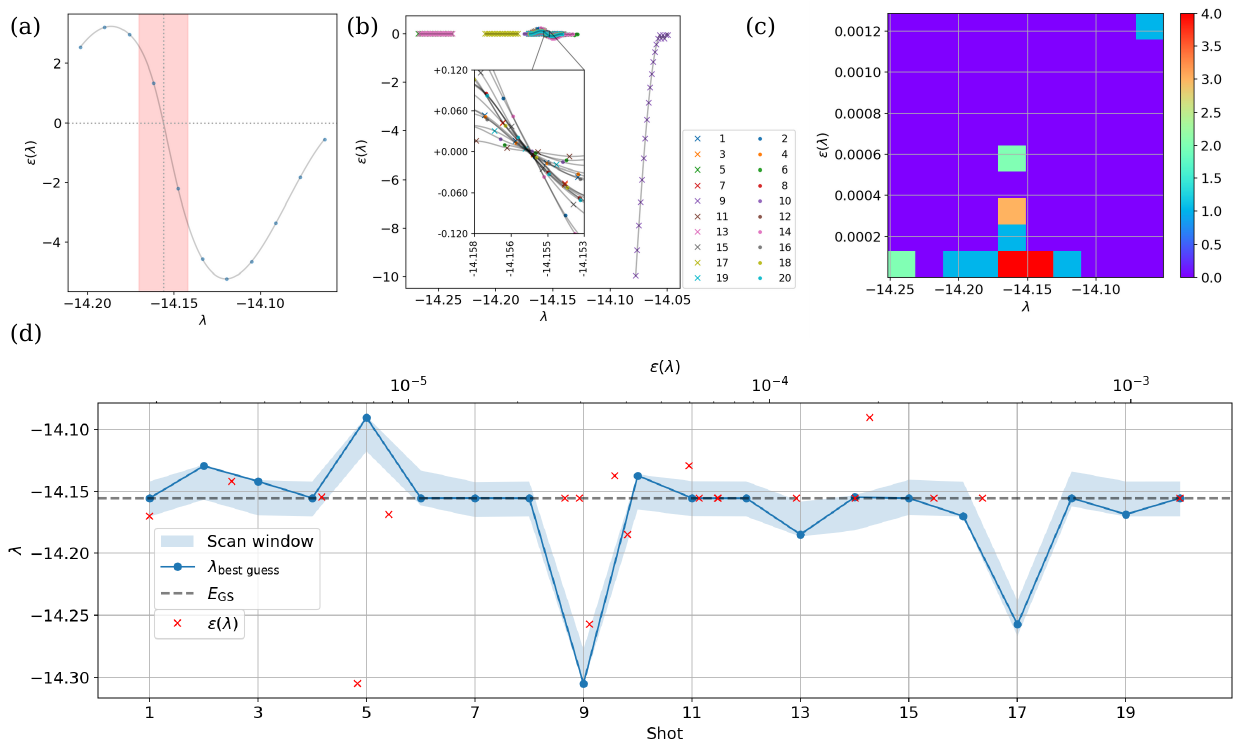}
    \caption{ (a) A coarse scan of the functional $\epsilon(\lambda)$ with simulated annealing around
    $H_{11}$ is performed to obtain an initial estimate of the Lagrange
    multiplier $\lambda$, shown by the grey curve. The interpolated root defines
    the center of the shaded pink region, which serves as the refined search
    interval for the next iteration. (b) Successive scans (labelled by the scan number in the legend) over progressively
    narrower intervals of $\lambda$ produce increasingly accurate estimates of
    the Lagrange multiplier. The inset shows the majority of the zero-crossings around the actual ground state eigenvalue. (c)
    Two-dimensional histogram of the roots in the scans of $\epsilon(\lambda)$; the highest count for $\lambda$
(red region) along with the widest range for $\epsilon(\lambda)$ gives the best guess range for $\lambda_\text{opt}$. (d) The progress of the successive scans shows the variation in width of the scanning window (blue shaded region) along with the estimated $\lambda_\text{opt}$ (blue points). The functional remains largely below the desired accuracy of 10 qubits and clustered around the actual $\lambda_\text{opt}$.}
    \label{fig:estimatelambda}
\end{figure*}

The Letter is organized as follows. The theory of the E1 transition rate is first
outlined, followed by an overview of the computational workflow. Next, our
modified QAE methodology for computing the ground states for different
angular-symmetry sectors is described, from which the E1 transition rates are
extracted. The results obtained from the D-Wave quantum annealer and their
comparison with GRASP
{are subsequently presented.}
Finally, we summarize our findings and discuss future directions.

We start our calculation with the Dirac--Coulomb Hamiltonian ($H_{\mathrm{DC}}$),
given by
\begin{equation}
H_{\text{DC}} =  \sum_i \left [c \bm{\alpha}_i \cdot \bm{p}_i+(\bm{\beta}_i-1)c^2+V_n(\bm{r}_i)\right]
 + \sum_{i > j}\frac{1}{r_{ij}} , \label{eq:HDC}
\end{equation}
where $\bm{p}_i$ denotes the momentum of the $i$-th electron, $\bm{\alpha}_{i}$
and $\bm{\beta}_{i}$ represent the Dirac matrix spinors, $c$ is the speed of
light, and $V_{\mathrm{n}}(\bm{r}_i)$ and $r_{ij}$ are the potential due to the
nucleus and the interelectronic separation, respectively.

A differential-equation-based MCDHF framework \cite{atoms11010007} is used to
generate a single-particle orbital basis, yielding a reduced number of highly
optimized orbitals and thus a more compact $H_{\mathrm{DC}}$ matrix and a smaller
set of expansion coefficients. Using this basis, the final wave functions for the
targeted states are obtained from a relativistic CI calculation based on the
$H_{\mathrm{DC}}$ Hamiltonian. The atomic state function (ASF),
$|\Psi(JM_J\Pi)\rangle$, for a state with total angular momentum $J$, parity
$\Pi$, and projection $M_J$, is expressed as a linear combination of
CSFs, $\Phi_r$:
\begin{equation}
|\Psi(JM_J\Pi)\rangle = {\sum_{r=1}^{B}} c_r |\Phi_r(JM_J\Pi)\rangle,
\label{asf}
\end{equation}
where $c_r$ are the expansion coefficients and $B$ is the number of CSFs.

The E1 transition rate $A$
\cite{atoms11010007}
can be extracted directly in terms of the ASF expansion from Eq.~(\ref{asf}) and
utilizing the Wigner-Eckart theorem.
\begin{equation}
A = \frac{2\omega}{3c} \frac{1}{(2J'+1)} \Big|\sum_{r,s} c_r^J c_s^{J'}\langle \Phi_r(J \Pi) \| \bm{D} \| \Phi_s(J' \Pi') \rangle\Big|^2
\label{avalue}
\end{equation}
where  $\langle \Phi_r(J \Pi) \| \bm{D} \| \Phi_s(J' \Pi') \rangle$ is the reduced
matrix element of the electric dipole transition operator $\bm{D}$
\cite{atoms11010007}.

We employed a hybrid quantum-classical framework that combines our modified QAE
algorithm with GRASP: the D-Wave quantum annealer determines the expansion
coefficients ($c_r^{J}$ and $ c_s^{J'}$) of ground and excited ASFs, while GRASP
supplies the $H_{\text{DC}}$ matrix elements and underlying reduced matrix
elements of the $\bm{D}$ operator. Next we compute $A$ by contracting the above
quantities according to Eq.~(\ref{avalue}). Our modified QAE framework is detailed
below.

\begin{table*}[t!]
\squeezetable
\centering
\caption{Correlation orbitals, excitation level, number of CSFs, and inactive core orbitals used in the GRASP calculations for
the ground (Gr.) and excited states (Ex.) of Li~I, Be~II, Na~I, and Mg~II. Here, the excitation type is denoted as (S = single, D = double) and in the fifth column, $P \rightarrow  Q$ means that the number of CSFs is truncated from $P$ to $Q$.   }
\label{tab:grasp-params}
\setlength{\tabcolsep}{12pt}
\begin{ruledtabular}
\begin{tabular}{llcccc}
\textbf{Species} & \textbf{State} & \textbf{Correlation} & \textbf{Excitation}
& \textbf{No.\ of} & \textbf{Inactive}\\
 & & \textbf{orbitals} & & \textbf{CSFs} & \textbf{core orbital}\\
\colrule
\multirow{2}{*}{Li~I}
  & Gr.\ ($^2S_{1/2}$) & $3s$ & SD & 8 & None\\
  & Ex.\ ($^2P_{1/2},{}^2P_{3/2}$) & $3p$ & SD & $15{\to}12,\ 21{\to}12$ & None\\
\colrule
\multirow{2}{*}{Be~II}
  & Gr.\ ($^2S_{1/2}$) & $3s$ & SD & 8 & None\\
  & Ex.\ ($^2P_{1/2},{}^2P_{3/2}$) & $3p$ & SD & $15{\to}10,\ 21{\to}12$ & None\\
\colrule
\multirow{2}{*}{Na~I}
  & Gr.\ ($^2S_{1/2}$) & $3p,4s$ & S & 11 & $1s$\\
  & Ex.\ ($^2P_{1/2},{}^2P_{3/2}$) & $3p,4s$ & S & $13{\to}10,\ 15{\to}12$ & $1s$\\
\colrule
\multirow{2}{*}{Mg~II}
  & Gr.\ ($^2S_{1/2}$) & $3p,4s$ & S & 11 & $1s$\\
  & Ex.\ ($^2P_{1/2},{}^2P_{3/2}$) & $3p,4s$ & S & $13{\to}10,\ 15{\to}11$ & $1s$\\
\end{tabular}
\end{ruledtabular}
\end{table*}

We start by defining the energy functional, as
\begin{equation}
    \epsilon(\lambda) = \braket{\Psi | H_{\text{DC}} | \Psi } - \lambda \braket{\Psi | \Psi }.
    \label{enfunc}
\end{equation}
where $\lambda$ is a Lagrange multiplier imposing the normalization constraint on the wave function. The ground state solution of Eq.(\ref{enfunc}) is to be found from the zero of the
energy functional and  $\ket{\Psi}$ is given in Eq. (\ref{asf}). Next, this
electronic structure eigenvalue problem is reformulated as a quadratic
unconstrained binary optimization (QUBO) problem, through a floating point
encoding scheme, suitable for  D-Wave's quantum annealing hardware, as
\begin{equation}
    \epsilon(\lambda) = \sum\limits_{r, s = 1}^{B} c_{r} c_{s} H_{rs}- \lambda \sum_{r=1}^{B} c_{r}^2,
\end{equation}
where $H_{rs} = \braket{\Phi_r | H_{\text{DC}} | \Phi_s }$.

{In the D-Wave annealer the lowest-energy eigenstate is found by
adiabatically evolving an initial transverse-field ground state into the final
QUBO ground state. Solving the QUBO problem yields a set of coefficients
$\{c_r\}$ in the ASF basis, and the optimal value of the Lagrange multiplier
$\lambda_{\text{opt}}$ corresponds to the ground-state energy of the system such that $\epsilon(\lambda_\text{opt})=0$.}

Within this encoding scheme, every expansion coefficient $c_{r}$ is represented by
a string of $K$ binary variables, in the manner set out in
\cite{kumar2024computation}. Thus a $B\times B$ Hamiltonian becomes a $BK\times BK$
QUBO matrix. Our modification of the QAE procedure introduces two further
ingredients for determining the optimal Lagrange multiplier
$\lambda_{\text{opt}}$ together with the coefficient set $\{c_{r}\}$, which we
refer to as \emph{zooming} and \emph{sigma-annealing}. We provide below an outline of these strategies, as described in \cite{paul2026applicationsquantumannealingmagnetic}, and test their applicability in this present work.

Here, sigma-annealing refers to the iterative convergence of {$c_r$} and
zooming proceeds as follows. We first scan $\lambda$ on a coarse grid in the
vicinity of $H_{11}$ (see Fig.~\ref{fig:estimatelambda}a) and interpolate the
resulting $\epsilon(\lambda)$ to locate its zero. The scan is repeated over a
slightly broader or narrower (randomly decided at each iteraton) window centered on that zero, and the procedure iterated. Successive
windows return an increasingly large fraction of roots clustered about the true
$\lambda_{\text{opt}}$, with correspondingly improving precision
(see Fig.~\ref{fig:estimatelambda}b). For better visualization, a 2D histogram is
shown in Fig.~\ref{fig:estimatelambda}c, where the roots are most densely
concentrated in the red region. The range best guess of $\lambda_\text{opt}$ is dictated by the higest count of roots and the widest range of $\epsilon(\lambda_\text{opt})$. We note that $\epsilon(\lambda_\text{opt})$ can have a spread in its values, therefore $\lambda \in [\lambda_\text{opt}-\delta\lambda_\text{opt}, \lambda_\text{opt}+\delta\lambda_\text{opt}]$ gives the highest count of solutions for $\epsilon(\lambda)=0$. Fig. \ref{fig:estimatelambda}d shows the progress of the successive scans with varying window-width (shaded blue region) and the estimated $\lambda_\text{opt}$. A variation (slight random increase or decrease) of the scanning window width and recentering occasionally causes a significant shift in the root of $\epsilon(\lambda)$ as seen in the three scans (shot numbers 5, 9 and 17) where the scan window misses the actual $\lambda_\text{opt}$, also corroborated by panel (b) of Fig. \ref{fig:estimatelambda}. We note that $\epsilon(\lambda)$ almost always remains below the target precision of 10 qubits, confirming the convergence of the algorithm. We find inclusion of zooming strategy to be
essential: without it neither simulated annealing (SA) \cite{doi:10.1126/science.220.4598.671} nor QA converges reliably
on $\lambda_{\text{opt}}$. The final and narrowest window returned by SA is then
passed to the D-Wave annealer. The plots in Fig.~\ref{fig:estimatelambda} are
obtained from multiple SA runs for the Be~II ($^2P_{3/2}$) case.

{Our quantum computations were performed on the D-Wave Advantage
system 6.4 quantum processing unit (QPU), based on the Pegasus topology
\cite{PhysRevApplied.21.034023, boothby2020nextgenerationtopologydwavequantum}
with over 5000 qubits, using 1000 reads and the default chain-coupling strength
\cite{dwave_chain_strength_2022} through D-Wave's Ocean SDK
\cite{ocean_sdk_2023}. We set $K=10$, which fixes the precision of the annealer
coefficients to three decimal places, and kept $B=11$ when working with the full
Hamiltonian, since the embedded graph on the D-Wave hardware is not all-to-all
connected. Where the CSF count exceeded 11 the basis was truncated: it was first
reordered using the independent-electron-pair approximation
$c_r \approx H_{r1}/(H_{11}-H_{rr})$, which gives a guess of the magnitude for
each CSF coefficient. Then rows and columns of $H_{\text{DC}}$ belonging to
negligible $c_r$ were dropped. This procedure worked for every atomic system
we studied.}

{Table~\ref{tab:grasp-params} lists the correlation-orbital sets, CSF counts, and
inactive-core choices used for the ground and lowest excited states. For the Li~I and 
Be~II systems, the full $1s^{2}\,n\ell$ configuration was left active and single
plus double (SD) excitations into a single correlating layer were admitted
($3s$ for the $^2S_{1/2}$ ground state, $3p$ for the $^2P_{1/2,3/2}$ excited
states). Because the $1s^{2}$ core is compact and the valence electron only
weakly screened, double excitations are the minimum required to capture the
leading pair correlation, which single excitations cannot describe. The resulting
spaces are small: $8$ CSFs for the ground state, and $15$ and $21$ CSFs for the
$J=1/2$ and $J=3/2$ odd states respectively. For the Na I and  Mg II systems, the $1s$ shell was frozen as an inactive core and
the expansion restricted to single (S) excitations into a $\{3p,4s\}$ correlating
set. This reflects a practical compromise: the $[\mathrm{Ne}]$-like core is large
enough that an unrestricted SD expansion would inflate the CSF count well beyond
what can currently be embedded on the annealer hardware, while single excitations
still recover the dominant contribution to the E1 transition rate.}

\begin{table*}[t!]
\squeezetable
\centering
\caption{GRASP transition energies and comparison of electric-dipole (E1)
transition rates $A$ (in s$^{-1}$) obtained from GRASP and D-Wave for Li~I,
Be~II, Na~I, and Mg~II. Transition energies are given in cm$^{-1}$. Percentage
errors of the D-Wave results are evaluated with respect to the GRASP reference
values and are defined as
$|(A_{\text{D-Wave}}-A_{\text{GRASP}})/A_{\text{GRASP}}|\times100\%$. The values of $A$ are presented to three decimal places corresponding to $K=10$ qubits in the QUBO formulation.}
\label{tab:e1-transitions}
\setlength{\tabcolsep}{14pt}
\begin{ruledtabular}
\begin{tabular}{llrrrr}
\textbf{Species} & \textbf{Transition} & \textbf{Transition energy}
& \textbf{GRASP $A$} & \textbf{D-Wave $A$} & \textbf{\% Error}\\
 & & \textbf{(cm$^{-1}$)} & \textbf{(s$^{-1}$)} & \textbf{(s$^{-1}$)} & \\
\colrule
\multirow{2}{*}{Li~I}
  & $^2S_{1/2}\rightarrow{}^2P_{1/2}$ & 13370.01 & $2.731\times10^{7}$ & $2.731\times10^{7}$ & $0.000$\\
  & $^2S_{1/2}\rightarrow{}^2P_{3/2}$ & 13370.57 & $2.731\times10^{7}$ & $2.731\times10^{7}$ & $0.000$\\
\colrule
\multirow{2}{*}{Be~II}
  & $^2S_{1/2}\rightarrow{}^2P_{1/2}$ & 30508.40 & $1.008\times10^{8}$ & $1.008\times10^{8}$ & $0.000$\\
  & $^2S_{1/2}\rightarrow{}^2P_{3/2}$ & 30517.33 & $1.009\times10^{8}$ & $1.009\times10^{8}$ & $0.000$\\
\colrule
\multirow{2}{*}{Na~I}
  & $^2S_{1/2}\rightarrow{}^2P_{1/2}$ & 15979.72 & $5.618\times10^{7}$ & $5.618\times10^{7}$ & $0.000$\\
  & $^2S_{1/2}\rightarrow{}^2P_{3/2}$ & 15995.98 & $5.634\times10^{7}$ & $5.634\times10^{7}$ & $0.000$\\
\colrule
\multirow{2}{*}{Mg~II}
  & $^2S_{1/2}\rightarrow{}^2P_{1/2}$ & 34651.67 & $2.546\times10^{8}$ & $2.546\times10^{8}$ & $0.000$\\
  & $^2S_{1/2}\rightarrow{}^2P_{3/2}$ & 34742.25 & $2.567\times10^{8}$ & $2.567\times10^{8}$ & $0.000$\\
\end{tabular}
\end{ruledtabular}
\end{table*}

We compare E1 transition rates, $A$, obtained from the annealer-derived coefficients (D-Wave) with those calculated using GRASP in Table~\ref{tab:e1-transitions}. As can be seen in Table~\ref{tab:e1-transitions}, the two remain identical to within the chosen numerical precision of three decimal places (corresponding to $K=10$ qubits). Since the D-Wave rates were evaluated using the transition energies obtained from the same GRASP orbital set, the entire deviation resides in the CI expansion coefficients $c_r^{J}$ and $c_s^{J'}$. The close agreement between the classical and annealer results therefore reflects the fidelity of the eigenvector estimation itself, and is consistent to three decimal places for which the coefficients are resolved. We attribute it to the iterative zooming and sigma-annealing strategy in our modified QAE, applied to both the ground and the excited state, together with a choice of correlating orbitals guided by the physics governing the E1 amplitude.

In summary, we have computed the E1 transition rates for Li~I, Be~II, Na~I, and
Mg~II using the modified QAE algorithm together with the CI method built on MCDHF
orbitals in GRASP. Because the annealer returns only the lowest state of a given
symmetry, the even- and odd-parity sectors were treated as separate problems. The
correlated orbitals were chosen to act on the valence dipole amplitude, and
combined with the iterative zooming and sigma-annealing strategy this gave
expansion coefficients and transition rates that reproduce the classical CI
values across all four systems. 

The framework carries over directly to other atomic properties. Polarizabilities
and isotope-shift parameters can also be evaluated from the same eigenvectors,
without repeating the anneal. On the hardware side, the D-Wave Advantage2 QPU
with its Zephyr topology~\cite{Boothby2022Zephyr} and the newer hybrid
solvers~\cite{osaba2024dwavesnonlinearprogramhybridsolver} should reduce
embedding overheads and improve solution quality. Together with better
preprocessing and error
mitigation~\cite{yip2025variationalquantumannealingquantum, Raymond2025,
Djidjev2024ReplicationBasedQA}, such developments may bring quantum annealers
into routine use for the accurate determination of atomic and molecular
properties.

\textit{Acknowledgment:} BP thanks Dr. Ranjan Modak (IIT Tirupati) for his
continuous support throughout this project. SD and BP thank Pradyot Pritam Sahoo
and Dr. Arup Chakraborty for useful discussions. PJ acknowledges support from the
Swedish Research Council (VR 2023-05367).\\
\textit{Data
{Availability}
Statement:} All data that support the findings of this study are included within the article and supplementary files.
\bibliography{t_rate}
\end{document}


\title{Supplemental
 Material}
\maketitle

\newpage 
\listoftables
\newpage 

\section{Ground State CSF coefficient tables}
The CSF coefficients are computed for ground states of different systems using GRASP and QA on D-Wave's annealing hardware, as listed in the following tables.



\begin{table}[h!]
    \centering
    \begin{tabular}{ccc}\hline
         CSF index&  GRASP& QA\\\hline
         1&  -9.99327E-01& 9.99E-01\\
         2&  -3.09311E-04& -3.48E-02\\
         3&  -9.24112E-04& 8.84E-03\\
         4&  2.63982E-03& 6.55E-03\\
         5&  -1.53516E-03& -2.56E-03\\
         6&  -6.56503E-03& 1.59E-03\\
         7&  -8.85121E-03& 9.06E-04\\
         8&  3.48274E-02& 2.76E-04\\\hline
    \end{tabular}
    \caption{CSF coefficients for Li ($^2S_{1/2}$) from a single- and double-excitation expansion with the $3s$ correlation orbital.}
    \label{tab:csfcoeffs_Li_3s_noSA}
\end{table}

\begin{table}[h!]
    \centering
    \begin{tabular}{ccc}\hline
         CSF index&  GRASP& QA\\\hline
         1&  9.99668E-01& 1.00E+00\\
         2&  2.42124E-04& -2.42E-02\\
         3&  -2.55474E-04& 7.59E-03\\
         4&  -3.43566E-03& 6.77E-03\\
         5&  1.59038E-03& -3.22E-03\\
         6&  6.18062E-03& 2.24E-03\\
         7&  7.98807E-03& 2.25E-05\\
         8&  -2.34018E-02& 9.46E-04\\\hline
    \end{tabular}
    \caption{CSF expansion coefficients for Be$^{+}$($^2S_{1/2}$), obtained from single and double excitations into the $3s$ correlation orbital.}
    \label{tab:csfcoeffs_Be_3s_noSA}
\end{table}

\begin{table}[h!]
    \centering
    \begin{tabular}{ccc}\hline
         CSF index&  GRASP& QA\\\hline
         1&  9.99942E-01& 1.00E+00\\
         2& -1.47100E-04& 7.17E-03\\
         3& -4.63802E-03& -4.64E-03\\
         4&  3.48711E-04& -4.57E-03\\
         5&  3.88697E-03& 3.83E-03\\
         6&  7.27083E-03& 1.76E-03\\
         7&  7.90288E-04& 7.69E-04\\
         8& -4.71327E-03& -6.32E-04\\
         9& 1.05072E-04& 3.43E-04\\
         10& 1.76770E-03& -1.48E-04\\
         11& -6.34766E-04& 1.07E-04\\\hline
    \end{tabular}
    \caption{CSF coefficients for Na ($^2S_{1/2}$) with 3p and 4s as correlation orbitals, single excitations and inactive 1s orbital.}
    \label{tab:csfcoeffs_Na_3p4s_S_in_1s_noSA}
\end{table}

\begin{table}[h!]
    \centering
    \begin{tabular}{ccc}\hline
         CSF index&  GRASP& QA\\\hline
         1&  -9.99940E-01& 1.00E+00\\
         2& 1.87040E-04& 7.02E-03\\
         3& 4.68714E-03& -4.75E-03\\
         4&  -1.74490E-04& -4.68E-03\\
         5&  -3.62178E-03& 3.62E-03\\
         6&  -7.04531E-03& 3.24E-03\\
         7& -3.94901E-04& -9.16E-04\\
         8& 4.77297E-03& 3.88E-04\\
         9& -7.91008E-05& -1.98E-04\\
         10& -3.23978E-03& 1.77E-04\\
         11& 9.22300E-04& 8.29E-05\\\hline
    \end{tabular}
    \caption{CSF coefficients for Mg$^{+}$ ($^2S_{1/2}$) with 3p, 4s as correlation orbital, single excitations and inactive 1s orbital.}
    \label{tab:csfcoeffs_Mg_3p4s_S_in_1s_noSA}
\end{table}

\section{Excited State CSF coefficient tables}
The CSF coefficients for excited states of different systems using GRASP and QA on D-Wave's annealing hardware are listed in the following tables. In cases where a truncated basis is used, the discarded CSFs are left blank in the QA column, and coefficients printed as $0.000000$ under GRASP have magnitudes below $10^{-6}$.

\begin{table}[h!]
    \centering
    \begin{tabular}{ccc}\hline
         CSF index&  GRASP& QA\\\hline
         1&  9.99147E-01& -9.99E-01\\
         2&  -1.19000E-04& 1.12E-04\\
         3&  8.47000E-04& -9.49E-04\\
         4&  3.00000E-06& \\
         5&  1.40000E-05& -7.95E-05\\
         6&  -3.48700E-03& 3.53E-03\\
         7&  4.49800E-03& -4.50E-03\\
         8&  -0.000000& 1.71E-06\\
         9&  -2.83800E-03& 2.76E-03\\
         10&  9.00000E-06& \\
         11&  -1.80000E-05& 1.32E-04\\
         12&  -1.10000E-05& 5.34E-05\\
         13&  3.33040E-02& -3.34E-02\\
         14&  0.000000& \\
         15&  2.35430E-02& -2.37E-02\\\hline
    \end{tabular}
    \caption{CSF coefficients of Li~I ($^{2}P_{1/2}$) obtained with GRASP and with the D-wave calculation, using the $3p$ correlation orbital with single and double (SD) excitations and no inactive core orbital. For the QA approach, a truncated basis of 12 configurations was used in place of the full 15 basis configurations.}
    \label{tab:csfcoeffs_Li_2p_J1half}
\end{table}


\begin{table}[h!]
    \centering
    \begin{tabular}{ccc}\hline
         CSF index&  GRASP& QA\\\hline
         1&  9.99147E-01& -9.99E-01\\
         2&  -1.18000E-04& 1.43E-04\\
         3&  5.99000E-04& -5.38E-04\\
         4&  0.000000& \\
         5&  5.99000E-04& -5.71E-04\\
         6&  -2.00000E-03& 1.94E-03\\
         7&  -4.50000E-03& 4.46E-03\\
         8&  -2.00000E-06& \\
         9&  0.000000& \\
         10&  -3.00000E-06& \\
         11&  -2.46700E-03& 2.41E-03\\
         12&  -3.18100E-03& 3.21E-03\\
         13&  1.00000E-05& \\
         14&  3.33000E-02& -3.33E-02\\
         15&  -1.40000E-05& 6.91E-05\\
         16&  8.00000E-06& \\
         17&  5.00000E-06& \\
         18&  2.35480E-02& -2.36E-02\\
         19&  -7.00000E-06& \\
         20&  1.40000E-05& -5.93E-05\\
         21&  6.00000E-06& \\\hline
    \end{tabular}
    \caption{CSF coefficients of Li~I ($^{2}P_{3/2}$) obtained with GRASP and with the D-wave calculation, using the $3p$ correlation orbital with single and double (SD) excitations and no inactive core orbital . For the QA approach, a truncated basis of 12 configurations was used in place of the full 21 basis configurations.}
    \label{tab:csfcoeffs_Li_2p_J3half}
\end{table}

\begin{table}[h!]
    \centering
    \begin{tabular}{ccc}\hline
         CSF index&  GRASP& QA \\\hline
         1&  9.99538E-01& -1.00E+00\\
         2&  -1.29000E-04& 1.71E-04\\
         3&  2.16800E-03& -2.14E-03\\
         4&  6.00000E-06& \\
         5&  3.50000E-05& -1.92E-05\\
         6&  -4.72300E-03& 4.71E-03\\
         7&  6.08800E-03& -6.13E-03\\
         8&  0.000000& \\
         9&  -3.83300E-03& 3.85E-03\\
         10&  1.00000E-05& -1.08E-05\\
         11&  -2.20000E-05& \\
         12&  -1.50000E-05& \\
         13&  2.37320E-02& -2.37E-02\\
         14&  0.000000& \\
         15&  1.67720E-02& -1.67E-02\\\hline
    \end{tabular}
    \caption{CSF coefficients of Be~II ($^{2}P_{1/2}$) obtained with GRASP and with the D-wave calculation, using the $3p$ correlation orbital with single and double (SD) excitations and no inactive core orbital . For the QA approach, a truncated basis of 10 configurations was used in place of the full 15 basis configurations.}
    \label{tab:csfcoeffs_Beplus_2p_J1half}
\end{table}

\begin{table}[h!]
    \centering
    \begin{tabular}{ccc}\hline
         CSF index&  GRASP& QA\\\hline
         1&  9.99538E-01& 1.00E+00\\
         2&  -1.29000E-04& -2.77E-04\\
         3&  1.53300E-03& 1.41E-03\\
         4&  0.000000& \\
         5&  1.53300E-03& 1.62E-03\\
         6&  -2.69100E-03& -2.63E-03\\
         7&  -6.09300E-03& -6.17E-03\\
         8&  -4.00000E-06& \\
         9&  0.000000& \\
         10&  -6.00000E-06& \\
         11&  -3.34100E-03& -3.41E-03\\
         12&  -4.30800E-03& -4.17E-03\\
         13&  2.50000E-05& \\
         14&  2.37270E-02& 2.39E-02\\
         15&  -1.80000E-05& 7.74E-05\\
         16&  9.00000E-06& \\
         17&  5.00000E-06& \\
         18&  1.67790E-02& 1.66E-02\\
         19&  -7.00000E-06& \\
         20&  1.80000E-05& 1.00E-04\\
         21&  6.00000E-06& \\\hline
    \end{tabular}
    \caption{CSF coefficients of Be~II ($^{2}P_{3/2}$) obtained with GRASP and with the D-wave calculation, using the $3p$ correlation orbital with single and double (SD) excitations and no inactive core orbital . For the QA approach, a truncated basis of 12 configurations was used in place of the full 21 basis configurations.}
    \label{tab:csfcoeffs_Beplus_2p_J3half}
\end{table}

\begin{table}[h!]
    \centering
    \begin{tabular}{ccc}\hline
         CSF index&  GRASP& QA\\\hline
         1&  9.99999E-01& 1.00E+00\\
         2&  -5.74000E-04& -5.80E-04\\
         3&  0.000000& \\
         4&  -6.83000E-04& -6.83E-04\\
         5&  3.78000E-04& 3.80E-04\\
         6&  -8.60000E-04& -8.69E-04\\
         7&  -5.73000E-04& -5.74E-04\\
         8&  2.38000E-04& 2.42E-04\\
         9&  5.20000E-05& \\
         10&  8.60000E-05& 8.89E-05\\
         11&  -2.00000E-04& -1.97E-04\\
         12&  9.30000E-05& 9.47E-05\\
         13&  1.90000E-05& \\\hline
    \end{tabular}
    \caption{CSF coefficients of Na~I ($^{2}P_{1/2}$) obtained with GRASP and with the D-wave calculation, using the $3p$ and $4s$ correlation orbitals with single (S) excitations and $1s$ as the inactive core orbital . For the QA approach, a truncated basis of 10 configurations was used in place of the full 13 basis configurations.}
    \label{tab:csfcoeffs_Na_3p_J1half}
\end{table}

\begin{table}[h!]
    \centering
    \begin{tabular}{ccc}\hline
         CSF index&  GRASP& QA \\\hline
         1&  9.99999E-01& 1.00E+00\\
         2&  -2.63000E-04& -2.66E-04\\
         3&  6.86000E-04& 6.91E-04\\
         4&  2.28000E-04& 2.36E-04\\
         5&  -7.81000E-04& -7.86E-04\\
         6&  -7.10000E-04& -7.22E-04\\
         7&  3.28000E-04& 3.31E-04\\
         8&  1.91000E-04& 2.02E-04\\
         9&  -4.15000E-04& -4.07E-04\\
         10&  5.60000E-05& \\
         11&  -1.87000E-04& -1.91E-04\\
         12&  -1.72000E-04& -1.78E-04\\
         13&  -8.60000E-05& \\
         14&  1.47000E-04& 1.47E-04\\
         15&  -1.41000E-04& \\\hline
    \end{tabular}
    \caption{CSF coefficients of Na~I ($^{2}P_{3/2}$) obtained with GRASP and with the D-wave calculation, using the $3p$ and $4s$ correlation orbitals with single (S) excitations and $1s$ as the inactive core orbital . For the QA approach, a truncated basis of 12 configurations was used in place of the full 15 basis configurations.}
    \label{tab:csfcoeffs_Na_3p_J3half}
\end{table}

\begin{table}[h!]
    \centering
    \begin{tabular}{ccc}\hline
         CSF index&  GRASP& QA\\\hline
         1&  9.99997E-01& 1.00E+00\\
         2&  -1.29900E-03& -1.29E-03\\
         3&  0.000000& \\
         4&  -9.90000E-04& -9.88E-04\\
         5&  8.65000E-04& 8.67E-04\\
         6&  -1.27100E-03& -1.26E-03\\
         7&  -4.08000E-04& -4.03E-04\\
         8&  7.66000E-04& 7.63E-04\\
         9&  2.00000E-05& \\
         10&  2.77000E-04& 2.73E-04\\
         11&  -5.19000E-04& -5.17E-04\\
         12&  2.68000E-04& 2.69E-04\\
         13&  1.00000E-04& \\\hline
    \end{tabular}
    \caption{CSF coefficients of Mg~II ($^{2}P_{1/2}$) obtained with GRASP and with the D-wave calculation, using the $3p$ and $4s$ correlation orbitals with single (S) excitations and $1s$ as the inactive core orbital . For the QA approach, a truncated basis of 10 configurations was used in place of the full 13 basis configurations.}
    \label{tab:csfcoeffs_Mgplus_3p_J1half}
\end{table}

\begin{table}[h!]
    \centering
    \begin{tabular}{ccc}\hline
         CSF index&  GRASP& QA\\\hline
         1&  9.99997E-01& -1.00E+00\\
         2&  -6.00000E-04& 6.00E-04\\
         3&  7.76000E-04& -7.66E-04\\
         4&  5.21000E-04& -5.20E-04\\
         5&  -1.37100E-03& 1.36E-03\\
         6&  -7.28000E-04& 7.18E-04\\
         7&  7.49000E-04& -7.51E-04\\
         8&  4.37000E-04& -4.38E-04\\
         9&  -9.42000E-04& 9.39E-04\\
         10&  3.30000E-05& \\
         11&  -6.04000E-04& 6.17E-04\\
         12&  -5.53000E-04& 5.68E-04\\
         13&  -1.90000E-04& \\
         14&  4.25000E-04& \\
         15&  -3.65000E-04& \\\hline
    \end{tabular}
    \caption{CSF coefficients of Mg~II ($^{2}P_{3/2}$) obtained with GRASP and with the D-wave calculation, using the $3p$ and $4s$ correlation orbitals with single (S) excitations and $1s$ as the inactive core orbital . For the QA approach, a truncated basis of 11 configurations was used in place of the full 15 basis configurations.}
    \label{tab:csfcoeffs_Mgplus_3p_J3half}
\end{table}